\documentclass{aastex631}

\usepackage[inline]{enumitem}

\accepted{April 29, 2022}

\graphicspath{{./}{figures/}}
\usepackage{xcolor}

\begin{document}

 \title{On-sky reconstruction of Keck Primary Mirror Piston Offsets using a Zernike Wavefront Sensor}

\author[0000-0002-0786-7307]{Maaike A.M. van Kooten}
\affiliation{University of California Santa Cruz \\
1156 High St. \\
Santa Cruz, CA 95064, USA}

\altaffiliation{For coorespondance email: mvankoot@ucsc.edu}

\author[0000-0002-0696-1780]{Sam Ragland}
\affiliation{W.M. Keck Observatory}

\author[0000-0003-0054-2953]{Rebecca Jensen-Clem}
\affiliation{University of California Santa Cruz \\
1156 High St. \\
Santa Cruz, CA 95064, USA}

\author[0000-0002-6171-9081]{Yinzi Xin}
\affiliation{California Institute of Technology}

\author[0000-0001-8953-1008]{Jacques-Robert Delorme}
\affiliation{W.M. Keck Observatory}

\author[0000-0001-5299-6899]{J. Kent Wallace}
\affiliation{Jet Propulsion Laboratory}
\affiliation{California Institute of Technology}



\begin{abstract}
The next generation of large ground- and space-based optical telescopes will have segmented primary mirrors. Co-phasing the segments requires a sensitive wavefront sensor capable of measuring phase discontinuities. The Zernike wavefront sensor (ZWFS) is a passive wavefront sensor that has been demonstrated to sense segmented-mirror piston, tip, and tilt with picometer precision in laboratory settings. We present the first on-sky results of an adaptive optics fed ZWFS on a segmented aperture telescope, W.M. Keck Observatory's Keck II. Within the Keck Planet Imager and Characterizer (KPIC) light path, the ZWFS mask operates in the H-band using an InGaAs detector (CRED2). We piston segments of the primary mirror by a known amount and measure the mirror's shape using both the ZWFS and a phase retrieval method on data acquired with the facility infrared imager, NIRC2. In the latter case, we employ slightly defocused NIRC2 images and a modified Gerchberg-Saxton phase retrieval algorithm to estimate the applied wavefront error. We find good agreement when comparing the phase retrieval and ZWFS reconstructions, with average measurements of $408 \pm23$~nm and $394 \pm 46$~nm, respectively, for three segments pistoned by 400~nm of optical path difference (OPD). Applying various OPDs, we are limited to $\sim$100~nm OPD of applied piston due to our observations' insufficient averaging of adaptive optics residuals. We also present simulations of the ZWFS that help explain the systematic offset observed in the ZWFS reconstructed data. 
\end{abstract}


%
\section{Introduction}
Future large space- and ground-based optical/near-infrared telescopes will have segmented primary mirrors. These new telescopes will have to initially co-phase the primary mirror segments and make fine adjustments to the phasing in order to support the image quality and/or contrast required for science observations. The Zernike wavefront sensor (ZWFS) is a passive wavefront sensor that can sense phase discontinuities, making it an ideal candidate for co-phasing segmented mirrors. In addition to being a fine-phasing sensor for segmented primary mirrors, the ZWFS can calibrate algorithms and correct for drifts in a system, acting as a general-purpose, low-order wavefront sensor. With this ability, a ZWFS will be used on the Nancy Grace Roman Space telescope~\citep{Shi_2016} to sense aberrations such as tip, tilt, focus, and astigmatism~\citep{Ruane_2020} in the system. Analytical studies by ~\cite{Laginja_2021} indicate the tolerance on mirror piston for specific segments on future space telescopes with a coronagraph for direct imaging of exo-earths is 7~pm. Simulations and laboratory experiments have shown that such picometer-precision can be achieved with the ZWFS~\citep{Moore_2018, Steeves_2020}. Finally, for segmented space telescopes the long stroke ZWFS has been proposed by~\cite{Jackson_2016} to increase the dynamic range by a factor of 20, allowing it to have a larger capture range. For future segmented ground-based telescopes, re-coating of the segments will be happening on a continuous rotation, with daily co-phasing of the telescope necessary; \cite{Cheffot_2020} have proposed the ZWFS for this purpose. Using the ZWFS under seeing-limited conditions to co-phase a segmented pupil (by having a segmented deformable mirror to provide an analogue to a segmented primary mirror), the Zernike unit for segment phasing (ZEUS) team, shows a precision of 3~nm on a 10th magnitude star~\citep{Dohlen_2006, Surdej_2010} and 4~$\mu m$ for multi-wavelength setup~\citep{Vigan_2011}. Working with diffraction-limited light,~\cite{Vigan_2018} achieve on-sky nanometer-accuracy corrections of non-common path aberrations (NCPA) within the instrument using the ZWFS on VLT/SPHERE~\citep[i.e., ZELDA;][]{Diaye_2016}. More recently,~\cite{Vigan_2022} show two temporal regimes of NCPAs on SPHERE, work that is only possible using the ZWFS. Finally, all of this work has illustrated the potential of the ZWFS as a powerful wavefront sensor that can measure low-order phase aberrations without probing the electric field dynamically and allowing the ZWFS to be used in parallel with science observations. Yet, the ZWFS has not previously been demonstrated on a segmented telescope. 

A ZWFS has recently been installed at the W.M. Keck Observatory's Keck II telescope within the Keck adaptive optics (AO) system through the Keck Planet Imager and Characterizer~\citep[KPIC;][]{Mawet_2016,Pezzato_2019} project. KPIC itself consists of several hardware upgrades such as a suite of new vortex coronographs and a Lyot stop for NIRC2, a near-infrared pyramid wavefront sensor ~\citep[PyWFS;][]{Bond_2020}, and a Fiber Injection Unit FIU;~\citep[FIU;][]{Delorme_2021} which contains the ZWFS. Combined with the development of new algorithms such as the predictive wavefront control~\citep{Jensen_2019,Vankooten_2021}, KPIC deployment has pushed toward better contrast on Keck II, while testing and verifying the new technology for future telescopes. These upgrades to Keck II have been decreasing the residual wavefront error and delivering better contrast with NIRC2~\citep{Vankooten_2021}. One major source of phase aberrations that is not well understood is the impact of phasing errors associated with Keck's segmented primary mirror. The phasing errors of the primary mirror impact the coronagraph's inner working angle and are not correctable by the AO system. Ideally, the low-order aberrations on the primary mirror from segment misalignment would be directly measured and then corrected by moving the segments themselves as demonstrated in simulation by~\cite{Janin-Potiron_2017}. The ZWFS is the first crucial step towards such a system as it can monitor the primary mirror’s shape. While the ZWFS will help better understand Keck's segmented primary mirror, it will also serve as a critical testbed for future space and ground-based telescopes that will need a fine-phase sensor such as the ZWFS. 

This paper presents initial on-sky results with the ZWFS installed on Keck II. We passively monitor the primary mirror using the ZWFS after inducing piston offsets onto the mirror's segments. We also implement a phase retrieval method using NIRC2 images to validate our measurements~\citep{Ragland_2018}. We first introduce some fundamentals of the ZWFS in Section~\ref{sec:zwfs}, with an overview of the on-sky data presented in Section~\ref{sec:onsky}. To understand features of the on-sky data, we perform simulations of the system, which are described in Section~\ref{sec:simulations}. Finally, we present the results of our on-sky tests and simulations in Section~\ref{sec:results}, followed by our conclusions and recommendations for future work in  Section~\ref{sec:conclusions}.

\section{The Zernike Wavefront Sensor} \label{sec:zwfs}

The ZWFS, a phase-contrast method, was first applied to observe phase objects as intensity objects in biological contexts. This work was awarded a Nobel prize in 1953. Proposed as a wavefront sensor for AO, the ZWFS transforms phase errors into amplitude errors, which enables a direct measurement of the electric field phase. \cite{Vigan_2018} performed an on-sky demonstration of the ZWFS for correction of non-common-path aberrations (NCPA), achieving a factor of 3 and 10 improvement in contrast at 300~mas and 600~mas, respectively, when operating in closed-loop. Optically, the ZWFS is a focal-plane mask (FPM; Zernike mask) that introduces a phase shift, followed by a camera in the downstream pupil plane. The classic Zernike FPM applies a static phase of $\pi /2$ to the core of the point-spread function (PSF; core being $\sim1 \lambda/D$ where $\lambda$ is wavelength and $D$ is the telescope diameter). Efforts to improve the ZWFS allowing the amplitude and phase of the electric field to be determined include a dynamic phase-shifting ZWFS FPM~\citep{Wallace_2011} and the vector-ZWFS~\citep{doelman_2018} that makes use of liquid-crystal, direct-write technology. There have also been many proposed improvements to the reconstruction algorithm, including work by~\cite{Steeves_2020}. 

\begin{figure}
\centering
\includegraphics[width=\hsize]{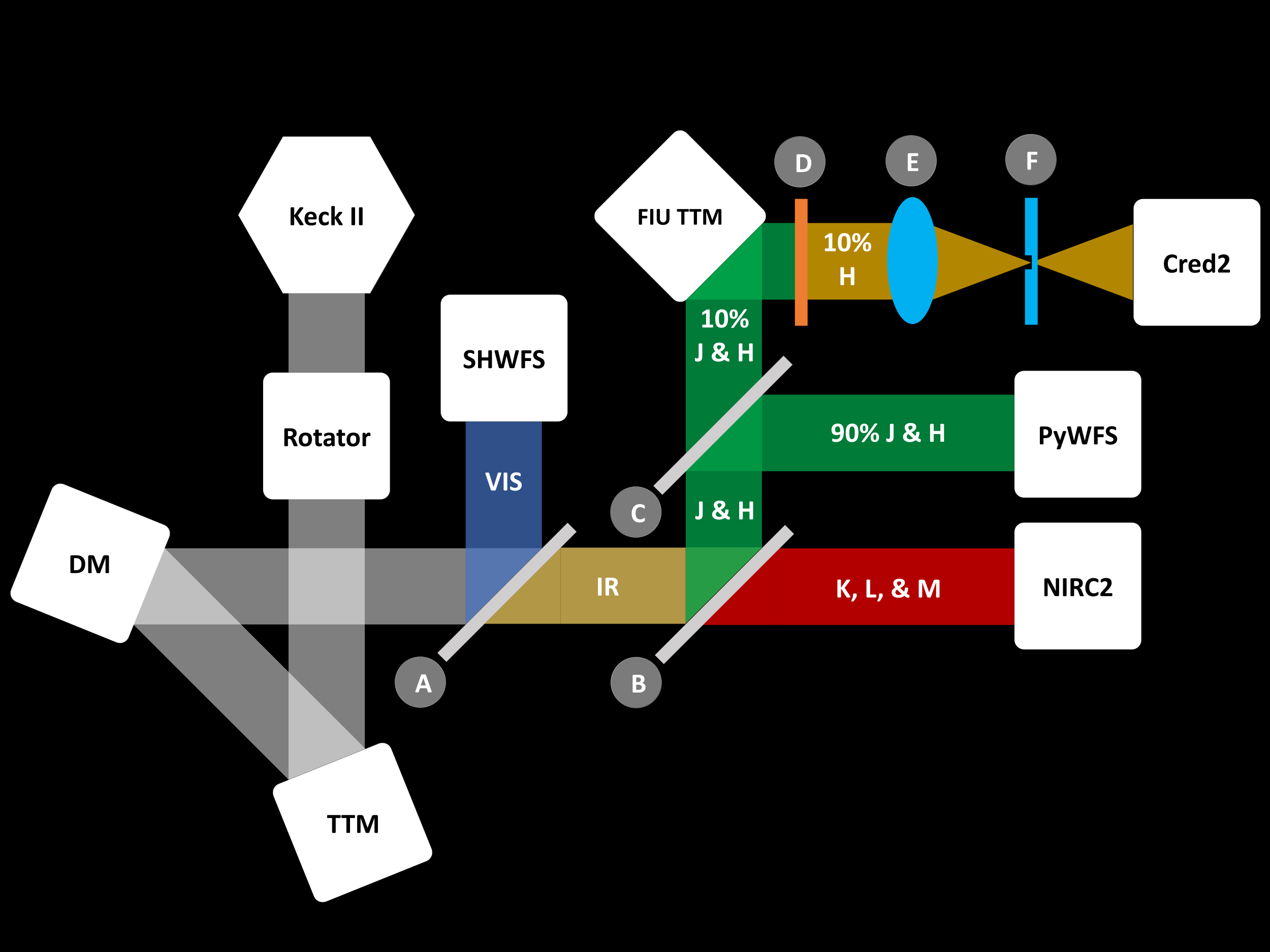}
    \caption{A schematic of the general light path in the Keck II AO bench and ZWFS path (also shared by KPIC). The TTM and DM correct the light from the telescope. The first dichroic (A) sends visible light to the SHWFS, and the KPIC pick-off arm (B) sends J and H band light KPIC and longer wavelengths to NIRC2. Within the KPIC path, 90\% of the light is sent to the PyWFS (C), while the remaining 10\% is sent to the FIU TTM. In our configuration, there is an H band filter (D) followed by a doublet lens (E) that focuses the light onto the Zernike mask (F). The Cred2 detector is placed in the pupil plane. }
    \label{fig:KeckIIAO}
\end{figure}

Figure~\ref{fig:KeckIIAO} presents a simplified schematic of the Keck II AO configuration used to acquire the data presented in this paper. The light coming from the telescope goes through the rotator before it is reflected by the facility tip-tilt mirror (TTM) and the deformable mirror (DM). The light is then split by the first dichroic (A), which sends the visible light to the Shack-Hartmann wavefront sensor (SHWFS) and transmits the infrared light to the KPIC pick-off (motorized stage holding a flat mirror and a dichroic; B). To acquire our data, we used the dichroic, which reflects the J and H bands toward KPIC and transmits longer light to NIRC2. The reflected J and H bands are split by the PyWFS pick-off (C), which reflects 90\% of J and H towards the PyWFS and transmits the rest of the light to the FIU TTM. This mirror folds the beam towards the optical assembly that contains the Zernike mask. The collimated beam goes through an H band filter (D) before it is focused on the FPM (F) by a lens doublet (E). After the FPM, the light diverges to form a pupil image ($\approx 400$ pixels in diameter) on a low noise InGaAs detector~\citep[\textit{First Light Imaging}, Cred2;][]{Gibson_2019}. 

The Zernike mask used to acquire the data presented in this paper was manufactured by \textit{Silios technology}. It consists of a 7.5~mm square optic cut into a 1-mm thick fused silica substrate. A circular dimple centered on the mask measures $8~\pm~2~\mu$m in diameter and is $790~\pm~10$~nm deep. The incoming beam onto the FPM has an f-number of 5.65. The edges of the substrate are coated with aluminum, and a crosshair is engraved into the substrate for alignment purposes. This classical Zernike mask applies a $\pi/2$ phase shift to the PSF core over $0.8788 \lambda/D$ with a top-hat profile. The mask was initially designed and optimized to operate in J and H bands (optimal wavelength of 1.4~$\mu$m); however, a static H band filter was later installed in front of the camera after the deployment of KPIC in order to reduce the impact of atmospheric dispersion when observing low-elevation targets with the KPIC FIU. Because the mask is not optimized for H band, the dimple diameter, corresponding to $0.88 \lambda/D$, limits its performance. The capture range of the ZWFS is fundamentally $\sim\lambda/4$ which is 398~nm for our setup. 
\section{On-sky data}\label{sec:onsky}
We took on-sky data using the ZWFS on the second half of the night of June 19, 2021, during engineering time. Observing bright targets, we applied segment piston-only patterns to the primary mirror with varying amplitudes. For each pattern on the mirror, we swapped between using the ZWFS to directly measure the phase aberrations on the primary mirror and NIRC2 where the phase retrieval algorithm~\citep{Ragland_2018} was applied to the images to estimate the phase. The AO system was configured to use the SHWFS for our observations. The SHWFS does not see the phase errors of the primary mirror since it is insensitive to piston and other modes (such as the segment tip-tilt) are below the noise floor of the SHWFS. 

\subsection{Zernike Wavefront Sensor}
For the ZWFS data acquisition we defined a dataset to consist of nine datacubes each containing 60 images of 0.5-s exposure, resulting in a total datacube exposure time of 30~s. The nine datacubes were taken at different alignment positions; since the ZWFS is very sensitive to the spot alignment on the FPM and on-sky alignment is a challenge, our strategy during observations was to scan a 3x3 grid of PSF positions using a steering mirror upstream of the Zernike mask and take data at each location. Note that an initial alignment was done during daytime tests. For each pattern, a reference dataset was taken to determine the nominal shape of the primary mirror. Then a segment offset pattern was applied and another dataset was acquired. We also acquired datasets without the mask by steering the PSF off of the Zernike mask.  

\begin{figure*}
   \resizebox{\hsize}{!}
            {\includegraphics[width=1.2\textwidth]{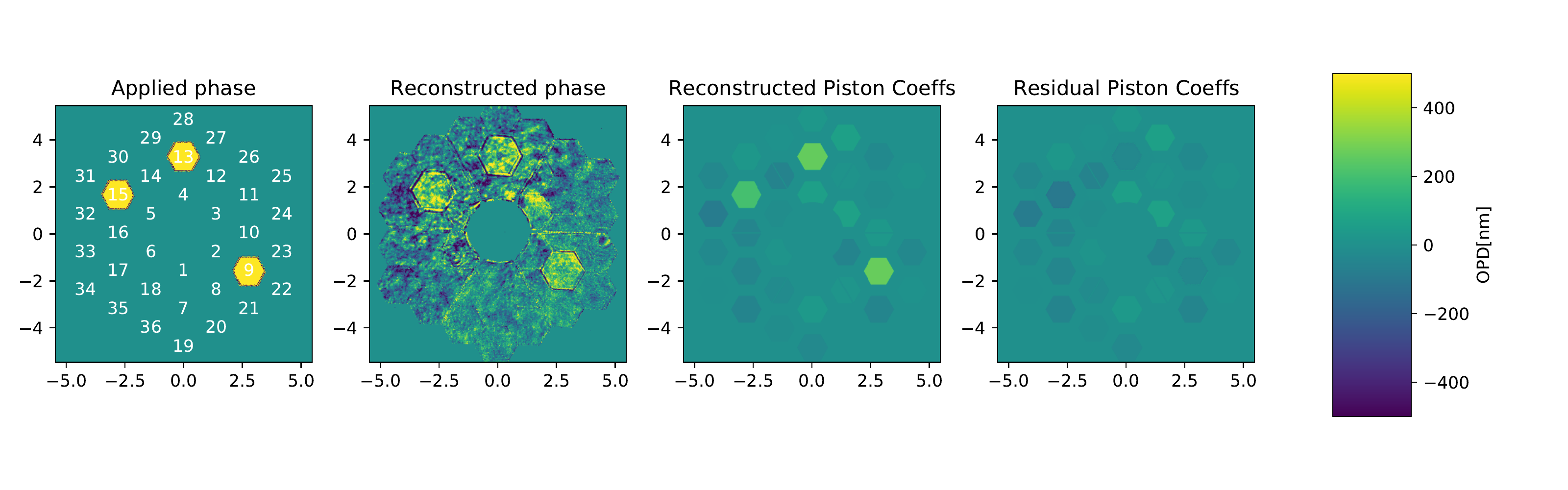}}
    \caption{Example of on-sky data taken with the ZWFS. First is the applied piston to the primary mirror with the actuator IDs labeled, second is the median combined reconstructed phase, third shows the piston coefficients projected onto the modal basis and the fourth is the residual piston values after subtracting off the applied piston values. The y- and x- axis are in units of $\lambda/D$.}
    \label{fig:example_data}
\end{figure*}

While we applied various patterns on the primary mirror, for the remainder of this paper we focus on an `L-shaped' pattern where three segments (segment numbers 9, 13, and 15) were pistoned by the same amount, probing two different spatial scales as shown in Figure~\ref{fig:example_data}. Observing V* BE Peg (M5 star with H-band magnitude of 2.32), we took data for various amounts of piston ranging from 25~nm to 200~nm surface error (i.e., 50~nm and 400~nm optical path difference; OPD). Note that 400~nm of OPD is at the upper limit of the ZWFS capture range. 

We performed the phase reconstruction using the algorithm outlined in~\cite{Diaye_2013} where the Strehl ratio (SR) is used to provide a more accurate reconstruction by estimating the apodization from the Zernike mask. We use this reconstructor since our applied phase shift of {$\pi/2$} is slightly smaller than the Airy disk in a top-hat profile and our amplitude function is pure transmission (fully transparent in the pupil and opaque outside the pupil as well as ignoring amplitude errors). Our Zernike mask sits behind an AO system allowing us to assume small phase errors, as required by the reconstructor. The phase, $\phi$, can then be reconstructed from the measured intensity, $I_c$

\begin{equation}
    \phi = -1+\sqrt{3-2b-(1-I_c)/b} ,
    \label{eq:recon}
\end{equation}
where $b$ is the apodization correction factor
\begin{equation}
    b=\sqrt{S}b_0 .
    \label{eq:b}
\end{equation}
$S$ is the SR and $b_0$ is the magnitude squared of the Zernike dimple with the input electric-field.

\begin{figure*}
 \centering
  \includegraphics[width=0.6\textwidth]{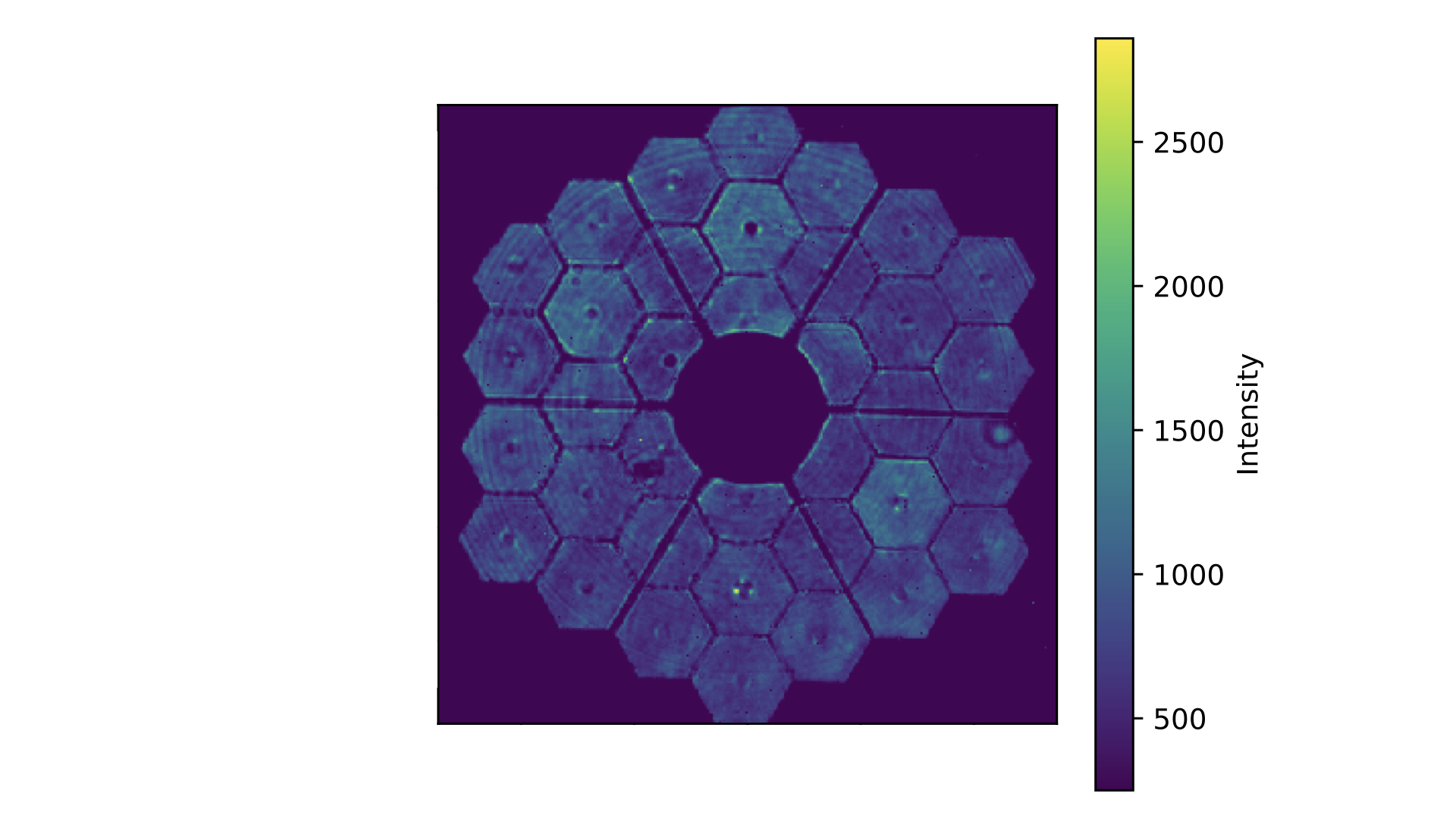}
    \caption{A raw ZWFS image with the Zernike mask in place. In the image the three segments are pistoned. The central dimple are a result of the bored support holes on the backside of the segments.}
    \label{fig:example_pupil}
\end{figure*}

\begin{figure*}
 \centering
            \includegraphics[width=0.6\textwidth]{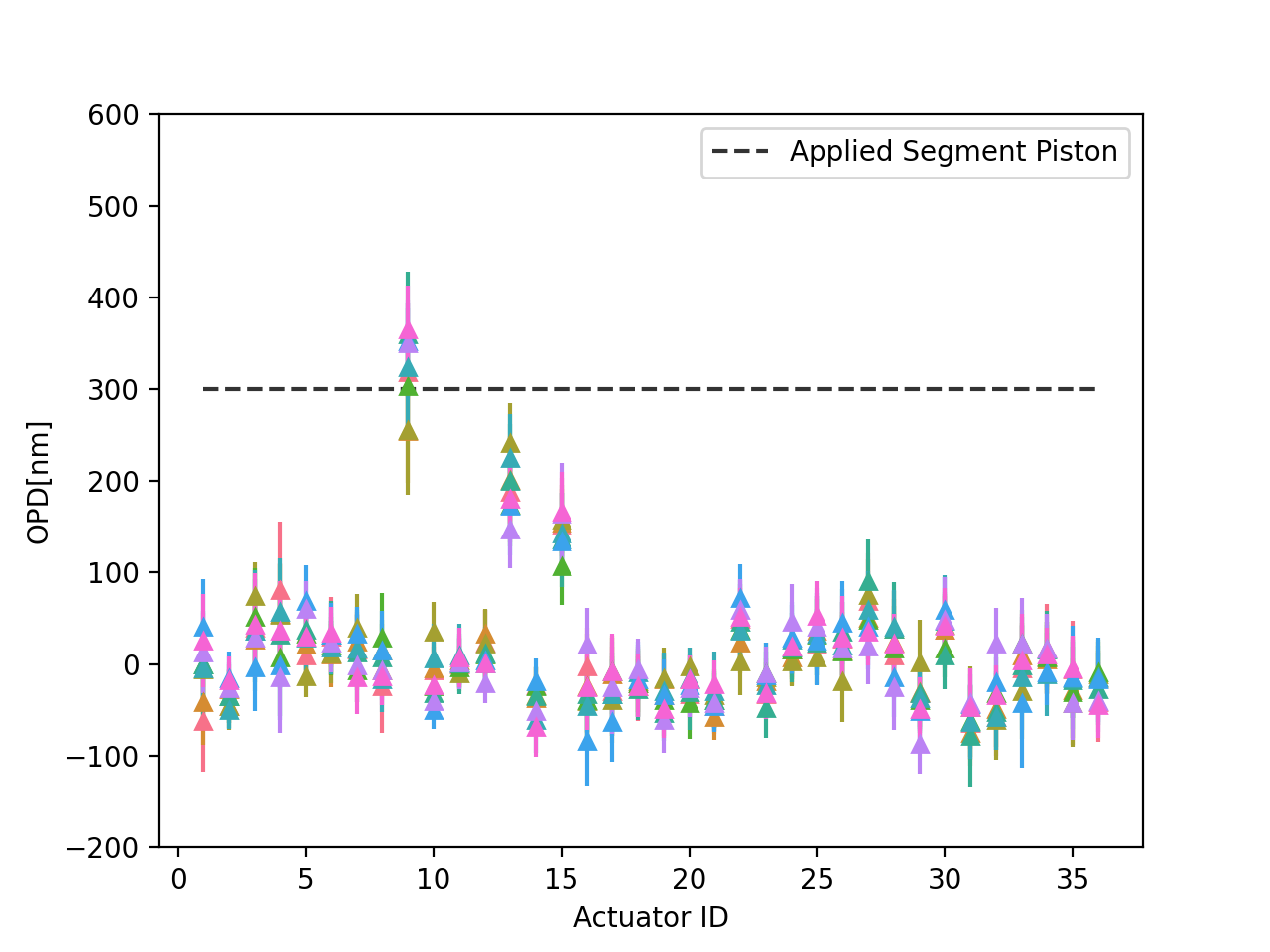}
    \caption{Reconstructed piston coefficients for each segment for the 9 different positions scanned with 300~nm segment piston applied. The 30-seconds mean and standard deviation of the reconstructed coefficients are shown.} 
    \label{fig:L_150nm_ZWFS}
\end{figure*}

We reconstruct the phase aberrations for each image in the dataset using the method above with $S=40~\%$. This value is on the lower end of typical H-band Strehl ratio values when NCPAs are accounted for; here, no NCPA calibration was done for the ZWFS. Note that with realistic Strehl ratios between 30\% and 70\% there is little change in the reconstructed values. The phase aberrations are then projected onto a segmented modal basis that reconstructs coefficients for mirror piston, tip, and tilt of each segment. We also follow these steps for the reference shape and then determine a mean reference shape from the 60 images in the datacube. The mean reference shape is modally subtracted from each frame in the other datacubes, and then a median for the coefficients for each position is determined. We then compare the segment piston coefficients to the amplitude of the applied piston. Figure~\ref{fig:L_150nm_ZWFS} shows the piston value for the 36 different segments of the Keck primary mirror compared to the applied piston value. One of the segments is in good agreement with the applied piston while two segments have slightly smaller reconstructed piston values. This indicates that either the sensitivity of the ZWFS varies across the pupil or that the primary mirror's segments respond differently to the same applied piston offset. Second, a sinusoidal pattern is apparent in the coefficients due to a tip-tilt across the entire pupil which is apparent in the ZWFS images. This signature in the figure is due to the segment numbering which spirals out from the center of the aperture. We investigate the source of the residual tip-tilt signal in Section~\ref{sec:simulations} through end-to-end simulations of the system. We also find evidence of slight misalignment by studying the intensity images of the ZWFS (i.e., second panel of Figure~\ref{fig:example_data}). We see that the thicknesses of the spider arms vary from one side of the aperture to another which is visible from close inspection of Figure~\ref{fig:example_pupil}, suggesting that the pupil plane is tilted relative to the camera focal plane.

\subsection{Phase Retrieval}
The observations for phase retrieval were taken with the NIRC2 Imager (the narrow camera; pixel scale: $\sim$10~mas) using the Br $\gamma$ filter (effective wavelength = 2.1685 $\mu$m) and the `open' NIRC2 pupil.  The setup involves defocusing the focus stage of the SHWFS by -5~mm to defocus (equivalent to 0.46 waves of defocus on NIRC2). Once again, we took two sets of measurements: (1) reference measurements without poking any primary mirror segment actuator, and (2) probe measurements by poking three segments (segment numbers 9, 13, and 15) by 400~nm OPD to form an L-shaped pattern. We collected 50 short exposures. Each exposure consists of 10 coadds with an integration time of 0.1 seconds per coadd resulting in each exposure having a total integration time of 1 second. We use coadding to improve the SNR in the NIRC2 exposure while reducing the overhead time as the stacking of the coadds is done onboard the detector.

An observational and modeling technique was proposed by co-author~\cite{Ragland_2018} to detect and account for possible systematic biases in phasing segments of a large-aperture optical telescope. The method cancels two bright speckles in the NIRC2 science images by moving the primary mirror segments of Keck. Here we use a modified Gerchberg-Saxton (MGS) algorithm~\citep{Gerchberg_1972} to estimate arbitrary segment piston errors~\citep{Ragland_2016} only. The results are presented in Section~\ref{sec:results} and compared to the ZWFS. Note, that the MGS method described here does not work well for smaller OPDs and hence only data was taken for 400~nm OPD.

\section{Zernike Wavefront Sensor Simulations}
\label{sec:simulations}
We simulate a ZWFS fed by an AO system matching our on-sky setup using the high contrast imaging package, HCIPy~\cite{hcipy}, in Python. We form a segmented primary mirror and use HCIPy's ability to have a deformable primary mirror through the segmented deformable mirror class. We implement Keck II's SHWFS with an F/15 incoming beam, 20 subapertures across the pupil, and a lenslet diameter of 0.2~mm. The SHWFS runs with a frame rate of 1200~Hz. A leaky integrator with a leak of 0.99 and a gain of 0.4 controls the 349-actuator DM. We use HCIPy's built-in ZWFS class, forming a ZWFS with a spot size of 0.88~$\lambda/D$. We set the ZWFS exposures, 0.5~s, to an integer number of the SHWFS frame rate to allow for averaging of the AO residuals and simulated 10~s of data. In total we acquire 20 frames which is one third the number of frames compared to on-sky data acquisition but to the computational load of simulating a full Keck AO system. During our observations, the CRED2 detector did not operate in non-destructive read mode; therefore we assume a read noise of 30 electrons and a dark current of 600 electrons per second per pixel in our simulation. We run single-layer turbulence with a Fried parameter of 20~cm, a wind speed of 5~m/s, and an outerscale of 50~m. For simplicity, both the SHWFS and ZWFS work at 1.59~$\mu$m.
The configuration of the simulations is summarized in Table~\ref{tab:simulation_parameters}.


\begin{table}
\centering
\begin{tabular}{||c c||} 
\hline
Parameter & Value \\[0.5ex] \hline \hline
$r_0$ & 20 cm \\ [0.5ex] \hline
$v$ & 5 m/s \\ [0.5ex]\hline
$L_0$ & 50 m \\ [0.5ex]\hline
Wavelength & $1.59 \mu m$\\ [0.5ex]\hline
SHWFS frame rate  & 1200 Hz\\ [0.5ex] \hline
SHWFS number subapertures & 20 \\ [0.5ex] \hline
SHWFS lenslet diameter & 0.2 mm\\ [0.5ex] \hline
DM number actuators & 21  \\ [0.5ex] \hline
Leak & 0.99 \\ [0.5ex] \hline
Gain & 0.4 \\ [0.5ex] \hline
Zernike spot size & 0.8788 $\lambda /D$ \\ [0.5ex] \hline
Zernike exposure time & 0.5 seconds \\ [0.5ex] \hline
Detector read noise & 30 $e^-$ \\ [0.5ex] \hline
Detector dark current & 600 $e^-$/s/pix \\ [0.5ex] \hline
Simulation length & 10 seconds \\ 
\hline
\end{tabular}
   \caption{Summary of the key parameters for our ZWFS simulations in HCIPy including the Fried parameter ($r_0$), wind speed ($v$), and outer scale ($L_0$) used to generate the single-layer atmosphere corrected by the SHWFS using a leaky integrator to control the DM. The ZWFS detector is the CRED2 detector that did not operate in non-destructive read mode during our observation.}
   \label{tab:simulation_parameters}
\end{table}

\begin{figure}
 \centering
        \includegraphics[width=\hsize]{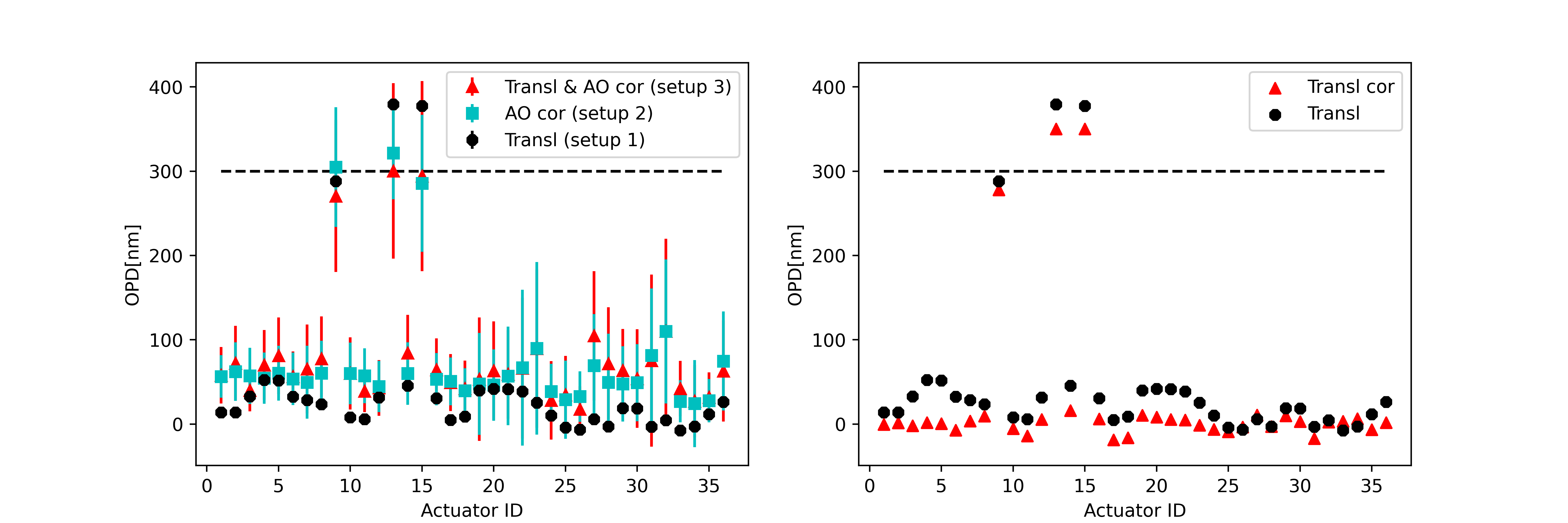}
    \caption{Left is the mean reconstructed coefficients and their standard deviation from simulated data with the three different simulation setups. Right shows the reconstructed coefficients for setup 1 showing the results before and after tip corrections. }
    \label{fig:Sim_ZWFS}
\end{figure}

In our on-sky data, we find a sinusoidal pattern in the reconstructed coefficients due to a tip-tilt across the entire pupil. Through simulations, we investigate the influence of a static translation of the ZWFS and the effects of residual atmospheric tip-tilt to determine what is responsible for the observed signature. We simulate three different setups: \begin{enumerate*}
    \item ZWFS with a translation of 0.005 $\lambda/D$,
    \item perfectly aligned ZWFS with atmospheric turbulence and AO correction, and
    \item ZWFS with a translation of 0.005 $\lambda/D$ and atmospheric turbulence with AO correction.
\end{enumerate*}
Using these simulations, we investigate the impact of a translation before the ZWFS mask or if the mask itself is tilted with respect to the incoming beam. In setup 2, we show the effect of only atmospheric tip-tilt on the data to understand how well the atmosphere is averaged during our on-sky observations. In the presence of residual atmospheric tip-tilt, the PSF location on the ZWFS mask moves around, effectively causing a small variable translation. We plot the results in Figure~\ref{fig:Sim_ZWFS}. The simulations with atmospheric turbulence have a larger spread in $1\sigma$ than the on-sky data shown in Figure~\ref{fig:L_150nm_ZWFS}. This difference is due to the simplifications used to model atmospheric conditions, the shorter length of the simulation (20 simulated frames verses 60 on-sky frames which means we expect 1.7 times worse $1\sigma$), and the fact we do not have a separate control loop for atmospheric tip-tilt. In the cases with a PSF translation, we see a pattern in coefficients of the unpoked segments similar to the pattern observed in the on-sky reconstruction. However, in Setup 2, with just atmospheric tip-tilt, we only find an increase in the $1\sigma$ and not a pattern. We, therefore, conclude that a static translation is the source for the observed pattern. By fitting a curve to the sinusoidal pattern of the coefficients, we can remove the effects of the translation, improving the reconstruction of the simulated data (right plot in Figure~\ref{fig:Sim_ZWFS}).

\section{Results}\label{sec:results}
\begin{figure*}
   \resizebox{\hsize}{!}
            {\includegraphics[width=\textwidth]{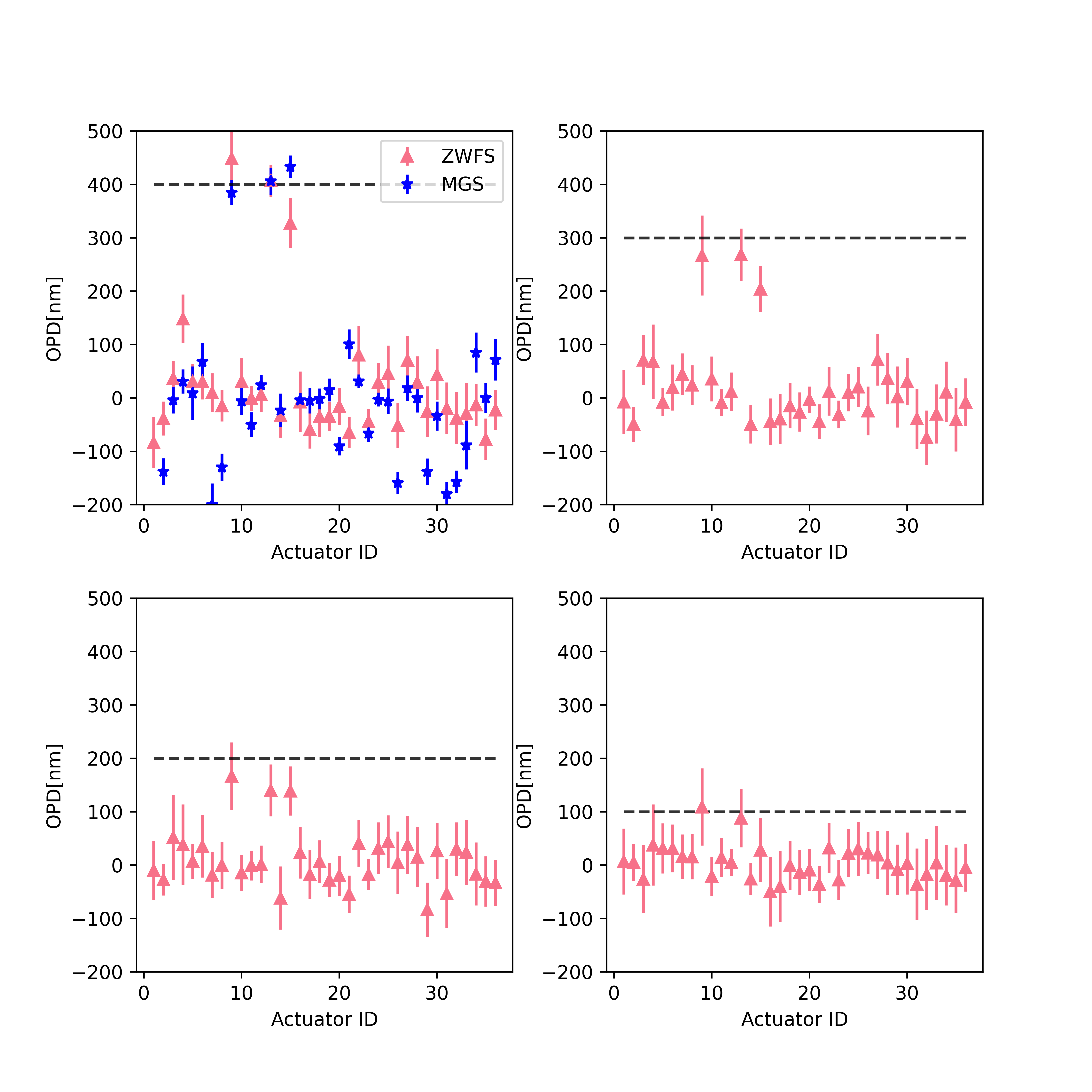}}
    \caption{The piston values for all 36 segments with varying amount of applied piston to three segments is indicated with the dashed black line. Top left has 400~nm of OPD applied, top right 300~nm, bottom left 200 nm, and bottom right 100 nm. Pink data points are ZWFS measurements while blue indicates the MGS data.  Residual tip-tilt across the primary mirror has been removed for each measurement (image) individually by fitting to the coefficients as described above. }
    \label{fig:L_ZWFS_all}
\end{figure*}

Through the simulations presented in Section~\ref{sec:simulations}, we show that a translation in the system can explain the difference in response to the applied piston for our three pistoned segments. We infer then that translation of the PSF relative to the FPM explains the sinusoidal pattern present in our on-sky data (e.g., Figure~\ref{fig:Sim_ZWFS}). We can subsequently improve our reconstruction of the on-sky data by removing the tip signal from the translation of the PSF for the cases with 400~nm, 300~nm, 200~nm, and 100~nm OPD, respectively (Figure~\ref{fig:L_ZWFS_all}). The MGS segment piston values with respect to the reference values for 36 segments are also presented in Figure~\ref{fig:L_ZWFS_all} for the 400~nm OPD case. The three poked segments are easily distinguishable. Using MGS, the $1\sigma$ scatter in the piston estimation for the 33 unpoked actuators is 90 nm. The three poked actuators give an average OPD of 408~nm with an average $1\sigma$ of 23~nm. For the ZWFS measurements, the $1\sigma$ of the unpoked actuators at 400~nm is 54 nm.  The ZWFS, close to its limits, found the average of the poked actuators to be 394~nm with a $1\sigma$ value of 46~nm. For the other datasets the ZWFS measured $247 \pm 56$~nm, $149 \pm 53$~nm, and $75 \pm 62$~nm for 300~nm, 200~nm, and 100~nm OPD, respectively. Finally, for the 200~nm OPD case, the unpoked $1\sigma$ is 32~nm and 24 nm for the 100~nm OPD case. For all the pokes, the applied piston is within one sigma of the value measured by the ZWFS. For the 100~nm OPD case, however, the $1\sigma$ is large, and we would be unable to distinguish all three pistoned segments from the noise of the unpoked segments. The results are summarized in Table~\ref{tab:summary_results}. 

The difference in scatter between all datasets relates to how well we are averaging out the atmospheric turbulence and the repeatability of the fast steering mirror moving the beam in the 3x3 grid. From the Mauna Kea Weather Center, the mean seeing value was 0.64" and 0.51" from the DIMM and MASS, respectively, for our given night. During our observations, the wind speed fluctuated between 3 and 6 m/s, suggesting a longer coherence time in the ground layer than typical (mean wind speed of 6.7 m/s; see KAON \#303). This means we need more exposures (or longer total integration time) to better average out the turbulence under these conditions. We also have residual tip-tilt between the reference shape and the pistoned shape due to the repeatability of the mirror. 

\begin{table}
\centering
\begin{tabular}{||c | c | c | c ||} 
\hline
Method & Applied segment piston [nm] & Measured segment piston [nm] & Non-piston segments [nm]\\[0.5ex] \hline \hline
MGS & 400 & 408 $\pm$ 23 & 90 \\
ZWFS & 400 & 394 $\pm$ 46 & 54 \\
ZWFS & 300 & 247 $\pm$53 & 37 \\
ZWFS & 200 & 149 $\pm$ 53 & 32\\
ZWFS & 100 & 75 $\pm$ 62 & 24 \\
\hline
\end{tabular}
   \caption{Summary of results for MGS and ZWFS showing the mean measured segment piston and mean $1\sigma$ values along with the mean $1\sigma$ values of the non-pistoned segments indicating the residual scatter.}
   \label{tab:summary_results}
\end{table}
\section{Conclusions and Future Work}\label{sec:conclusions}
We present on-sky results of the ZWFS where we piston three segments of Keck II's primary mirror with various amplitudes. For our 400~nm amplitude piston offset we reconstruct, from the ZWFS data, $394 \pm 46$nm OPD on average for the three segments. We compare this result to the MGS phase retrieval algorithm using NIRC2 images and find good agreement with the method, which estimates $408 \pm23$nm at 400~nm of the applied piston. This result confirms that the piston applied to the primary mirror in open-loop is consistent with our ZWFS reconstruction. We also show via simulations that a slight systematic difference in the reconstructed phase is due to an internal translation in the system, which must be taken into account in the reconstruction. Going to smaller OPD, we find that we cannot distinguish the poked segments for our 100~nm amplitude piston offset case due to the noise from the unpoked segments. This noise floor is not intrinsic to the ZWFS, but rather is driven mainly by our insufficient averaging out of the turbulence residuals during these observations. 

Future work will focus on improving the ZWFS sensitivity such that it will be able to measure and correct for smaller OPDs on the primary mirror. In early 2022, KPIC will be upgraded to include a better ZWFS as well as other modules that will improve the overall performance of the ZWFS. After the installation of the new ZWFS mask, we will work on improving the on-sky alignment of the ZWFS and optimizing the exposure time and the number of frames for the seeing and wind-speed conditions. We will investigate iterative reconstruction algorithms to reconstruct the phase from the ZWFS images better. Efforts to take passive measurements during science observations using the ZWFS are underway, where we hope to better understand how the primary mirror shape changes as a function of elevation.  Finally, we are working towards implementing a slow feedback loop between the ZWFS and the primary mirror such that we can maintain the best co-phasing of the segments throughout a science observation and demonstrate this capability for future telescopes, both ground- and space-based.
\section*{acknowledgments}

This work is funded by the Heising-Simons Foundation. The W. M. Keck Observatory is operated as a scientific partnership among the California Institute of Technology, the University of California, and the National Aeronautics and Space Administration. The Observatory was made possible by the generous financial support of the W. M. Keck Foundation. This research has made use of the Keck Observatory Archive (KOA), which is operated by the W. M. Keck Observatory and the NASA Exoplanet Science Institute (NExScI), under contract with the National Aeronautics and Space Administration. The authors wish to recognize and acknowledge the very significant cultural role and reverence that the summit of Mauna Kea has always had within the indigenous Hawaiian community. We are most fortunate to have the opportunity to conduct observations from this mountain. 

The work benefited from the NASA Nexus for Exoplanet System Science (NExSS) research coordination network sponsored by the NASA Science Mission Directorate. A portion of this research was also carried out at the Jet Propulsion Laboratory, California Institute of Technology under a contract with the National Aeronautics and Space Administration (NASA).

%
\bibliographystyle{aa} 
\bibliography{references} 
%

\end{document}